\documentclass[aps,pra,nofootinbib,preprint,amsmath,amssymb,floatfix]{revtex4-2}
\usepackage{color}
\usepackage{graphicx}

\begin{document}

\newcommand{\tc}{\textcolor}
\newcommand{\g}{blue}
\title{Axion electrodynamics: Energy-momentum tensor, and possibilities for  experimental tests}         
\author{ Iver H. Brevik$^1$  }      
\affiliation{$^1$Department of Energy and Process Engineering, Norwegian University of Science and Technology, N-7491 Trondheim, Norway}
\author{Moshe M.  Chaichian$^2$}
\affiliation{$^2$ Department of Physics, University of Helsinki, O.O.Box 64, FI-00014 Helsinki, Finland}

\date{\today}          

\begin{abstract}

Axion electrodynamics is based upon the Lagrangian of the electromagnetic (EM) field plus its interaction with the axions, and is accordingly a physically open system. It means that the four-divergence of the EM energy-momentum tensor is different from zero, implying in turn that the total EM energy and momentum (when integrated over all space) do not constitute a four-vector. The EM force is in principle accessible to experimental detection, just analogous to what is the case in ordinary electrodynamics. In the first half of this paper the energy-momentum aspects of axion electrodynamics are worked out in general, when the surroundings are allowed to be a medium with constant permittivity and permeability. In the second half, two examples are discussed. The first is a static situation, where a block of uniform  material containing axions is exposed to external strong electric and magnetic fields. Assuming the axion amplitude $a(x)$ (i.e. its density) to increase linearly in one direction, we calculate the axion-generated forces. As a second example, we consider axions varying not with position but instead harmonically  with  time; this is the constellation usually assumed in astrophysics. Assuming a Gaussian profile for the EM wave emitted from the Earth towards an axion cloud in outer space, we make a calculation of the 'axion echo', the return signal.
\end{abstract}
\maketitle

\bigskip
\section{Introduction}
\label{secintro}

Pseudoscalar axions  of amplitude $a= a(x)$ ($x$ being spacetime), which serve as  one of the leading candidates for dark matter in the Universe \cite{sikivie83,preskill83,abbott83,dine83},
arise as a consequence of the need for a strong CP-violation \cite{peccei77,peccei77a,weinberg78}.  If they turn out to be  detectable   experimentally, this would mean an important step  in our understanding of the Universe's composition and its evolution.  The axions are believed to be all-pervading, hardly interacting with ordinary matter at all. They are "cold", meaning that they are moving with nonrelativistic velocity $v \sim 10^{-3}c$. The  axion mass $m_a$ is open to a wide range of possibilities; most commonly it is  assumed to be of the order of tens of  $\mu$eV/c$^2$. These particles may have been produced at a very early stage
 in the Universe's history, about the inflationary times. The existence of them was suggested by  Peccei and  Quinn in 1977 in connection with the strong CP violation \cite{peccei77},    and the subject has since  attracted considerable interest. Some  works on axion electrodynamics can be found in \cite{sikivie14,lawson19,asztalos04,kim19,qingdong19,sikivie03,mcdonald20,chaichian20,zyla20,arza20,carenza20,leroy20,brevik20,brevik21a,oullet19,arza19,
qiu17,dror21,millar17,brevik22a,fukushima19,tobar19,adshead20,bae22,liu22} and references therein.

Since the axions should be  present everywhere, it could be in principle possible to detect them in terrestrial experiments.  In astrophysical contexts, it is common to assume that they are spatially uniform, ${\bf \nabla}a=0$, but vary periodically in
  time with low frequency $\omega_a$. One popular  suggestion  about how to detect axions on the Earth is presented in Refs.~\cite{sikivie14,lawson19,asztalos04,kim19}, where the basic idea is to observe   the resonance between the natural electromagnetic oscillations in a long plasma cylinder and  the axion field oscillations. A strong magnetic field in the axial $z$ direction is then required, and   extra measures turn out to be  necessary  to get the cylinder   'dilute' enough to make  the electromagnetic resonance frequencies, of the order of 100 GHz,  observable.

  Our aim in the present paper, however, is not  to discuss experiments in detail, but instead to focus on the electromagnetic energy-momentum tensor for the coupled axion field. As the theory is derived from a Lagrangian containing the electromagnetic field plus its coupling to the axions, the energy-momentum  tensor will reflect  a physically open system and thus  the components of electromagnetic energy and momentum when integrated over all space will not constitute  a four-vector. This follows directly from the fact that the EM energy-momentum tensor has a four-divergence different from zero, and is a consequence of relativistic electrodynamics. These features are important also for the axionic extension of the theory.

  In the next section we present the basic formalism of axion electrodynamics,  allowing for a dielectric environment with constant permittivity and permeability. The covariant properties of macroscopic electrodynamics in media are thereby shown and the  electromagnetic energy-momentum  tensor is constructed, together with the components of the four-force density.

  In the second half of the paper we analyze two specific cases where the interaction between photons and axions  comes into play: (1) the axion-produced fields and the forces acting on   a dielectric slab exposed to strong static orthogonal electric and magnetic fields, and (2) the axion echo idea developed as a possibility for detecting axions in astrophysics. The new element in our last-mentioned  analysis is the adoption of a Gaussian beam from the Earth directed to the axion cloud.

\section{Basics of axion electrodynamics}

The fundamental process is the interaction between a pseudoscalar axion and two photons \cite{mcdonald20}. The Lagrangian describing the electromagnetic field in interaction with the axion field is
\begin{equation}
{\cal{L}}= -\frac{1}{4}F_{\alpha\beta}{H}^{\alpha\beta}  - \frac{1}{4}g_\gamma \frac{\alpha}{\pi}\frac{1}{f_a}a(x) F_{\alpha\beta}\tilde{F}^{\alpha\beta}. \label{1}
\end{equation}
Here $g_\gamma$ is a model-dependent constant of order unity; for definiteness  we adopt the value  $g_\gamma = 0.36$ which follows from the DFSZ model \cite{sikivie03,dine81}. Further, $\alpha$ is the usual fine structure constant, and $f_a$ is the axion decay constant whose value  is only
insufficiently known; it is often assumed that $ f_a   \sim 10^{12}~$GeV. We assume an isotropic and homogeneous dielectric background, with constant permittivity $\varepsilon$ and permeability $\mu$. When the medium is at rest, the constitutive relations are $\bf{D}=\varepsilon {\bf E}, \ {\bf B}=\mu \bf H$. As is known, there are two field tensors, the basic tensor $F_{\alpha\beta}$ and the dielectric response tensor $H_{\alpha\beta}$, where $\alpha$ and $\beta$ run from 0 to 3. We will use the metric convention $g_{00}= -1$. The second term in Eq.~(\ref{1}) should be a total derivative (topological invariant) when the axion $a=$const, what is obviously true for $F_{\alpha\beta}$ as second factor but not the case with $H_{\alpha\beta}$, for instance.

 The quantity  multiplying the axion field $a(x)$ in Eq.~(\ref{1}) is thus the product of the electromagnetic field tensor $F_{\alpha\beta}$ and the dual,   $\tilde{F}^{\alpha\beta}=\frac{1}{2}\varepsilon^{\alpha\beta\gamma\delta}F_{\gamma \delta}$, with $\varepsilon^{0123}= 1$. It is convenient to give the expressions for the field tensors explicitly,
\begin{equation}
F_{\alpha\beta}= \left( \begin{array}{rrrr}
0    &  -E_x   & -E_y  &  -E_z \\
E_x  &    ~0     & B_z   & -B_y  \\
E_y  &  -B_z   &  ~0    &  B_x \\
E_z  &   B_y   &  -B_x &   ~0
\end{array}
\right), \label{2}
\end{equation}
\begin{equation}
H^{\alpha \beta} = \left( \begin{array}{rrrr}
0    &  D_x    &  D_y    &  D_z  \\
-D_x &  0      &  H_z   &  -H_y  \\
-D_y &  -H_z    &   0     &  H_x  \\
-D_z &  H_y   &   -H_x   &    0
\end{array}
\right). \label{2a}
\end{equation}
The following relations are also useful,
\begin{equation}
 F_{\alpha\beta}H^{\alpha\beta}={ 2(\bf{H\cdot B-E\cdot D)}}, \quad  F_{\alpha\beta}\tilde{F}^{\alpha\beta}= -4\,{\bf E\cdot B}. \label{3}
 \end{equation}
We define the  combined axion-two-photon coupling constant  as
\begin{equation}
g_{a\gamma\gamma}= g_\gamma \frac{\alpha}{\pi}\frac{1}{f_a},
\end{equation}
so that  the last term in the Lagrangian (\ref{1}) can be written as
\begin{equation}
{\cal{L}}_{a\gamma\gamma} =  g_{a\gamma\gamma} a(x)\,{\bf E\cdot B}. \label{4a}
\end{equation}
Based on the  expression (\ref{1}), the extended Maxwell equations take the following form,
\begin{equation}
{\bf \nabla \cdot D}= \rho-g_{a\gamma\gamma}{\bf B\cdot \nabla}a, \label{5}
\end{equation}
\begin{equation}
{\bf \nabla \times H}= {\bf J}+\dot{\bf D}+g_{a\gamma\gamma}\dot{a}{\bf B}+g_{a\gamma\gamma}{
\bf \nabla}a\times {\bf E}, \label{7}
\end{equation}
\begin{equation}
{\bf \nabla \cdot B}=0, \label{8}
\end{equation}
\begin{equation}
{\bf \nabla \times E} = -\dot{\bf B}. \label{9}
\end{equation}
Here $(\rho, {\bf J})$ are the usual electromagnetic charge and current densities. The equations are  general;, i.e. there are    no restrictions so far  on the spacetime variation of $a(x)$. The equations are moreover relativistic covariant, with respect to shift of the inertial system.

The governing equations for the fields can correspondingly be written as
\begin{equation}
\nabla^2 {\bf E}-\varepsilon\mu \ddot{\bf E}=    {\bf \nabla (\nabla \cdot E)}
 +\mu \dot{\bf J}+ \mu g_{a\gamma\gamma}\frac{\partial}{\partial t}\left[\dot{a}{\bf B}+ {\bf \nabla}a{\bf \times E}\right], \label{10}
\end{equation}
\begin{equation}
\nabla^2 {\bf H}-\varepsilon\mu \ddot{\bf H}= -{\bf \nabla \times J}-g_{a\gamma\gamma}{\bf \nabla \times }[\dot{a}{\bf B}+{\bf \nabla}a{\bf \times E}]. \label{11}
\end{equation}
The following relation is also useful,
\begin{equation}
{\bf \nabla (\nabla \cdot E)}= \frac{1}{\varepsilon}{\bf\nabla}[ \rho- g_{a\gamma\gamma}{\bf B \cdot \nabla}a].
\end{equation}
Note that the dynamical field entities are here the electromagnetic fields. We limit ourselves to a perturbative approach, looking upon the axions as a perturbation.  We do not consider the field equations for the axions explicitly (as mentioned, the physical system is an open system), and thus the axions are given external fields in the considered theory.

An important point is that the field equations can be significantly simplified in practical cases, the   reasons being   that axions appear usually in connection with strong external fields, electric or magnetic, or in combination.  Let us assume that there is only one such field, of magnetic type, called  ${\bf B}_0$, directed along the $z$ axis. The total electric field $\bf E$ can be decomposed as ${\bf E}= {\bf E}_a + {\bf E}_\gamma$, where the first term is the axion-generated part and the second the photon part, associated with the electromagnetic wave. The second order derivative $\ddot{a}$ in Eq.~(\ref{10}) can be taken to be associated with ${\bf B}_0$, so that the axion-generated field obeys the reduced equation
\begin{equation}
\nabla^2 {\bf E}_a-\varepsilon\mu \ddot{\bf E}_a=    \mu g_{a\gamma\gamma}\ddot{a}{\bf B}_0, \label{ 14a}
\end{equation}
in which we have put $\rho$ and ${\bf \nabla}a$ equal to zero. With $a=a_0\sin \omega_a t$ this yields the solution
\begin{equation}
{\bf E}_a(t)= -\frac{1}{\varepsilon} g_{a\gamma\gamma}a(t)B_0 \,\hat{\bf z}, \label{15a}
\end{equation}
so the axion-generated electric field is parallel or antiparallel to the applied magnetic field. It is also convenient to introduce a separate symbol $E_0$ for the constant electric field
\begin{equation}
 E_0= g_{a\gamma\gamma} a_0 B_0. \label{definition}
 \end{equation}
Now having separated off the component ${\bf E}_a$, the field equation (\ref{10}) for the electromagnetic wave ${\bf E}_\gamma$ takes the reduced form
 \begin{equation}
\nabla^2 {\bf E}-\varepsilon\mu \ddot{\bf E}=  {\bf \nabla (\nabla \cdot E)} +\mu {\dot{ \bf J}}+  \mu g_{a\gamma\gamma}[ \dot{a}\dot{\bf B} + {\bf \nabla}a{\bf \times \dot{E}}], \label{12}
\end{equation}
(subscript gamma omitted). Likewise for the magnetic field
\begin{equation}
\nabla^2 {\bf H}-\varepsilon\mu \ddot{\bf H} = -{\bf \nabla \times J}-g_{a\gamma\gamma}\left[   \dot{a}{\bf \nabla \times B}
 + ({\bf \nabla}a){\bf \nabla \cdot E}
 -[{\bf \nabla}a \cdot {\bf \nabla}]  {\bf E}      \right]. \label{13}
\end{equation}
As a general remark, it ought to be recalled that in the electromagnetic theory of media  it is the magnetic  induction $\bf B$ that is the fundamental magnetic field  as it takes  into account the presence of the molecules directly. This is the reason why $\bf B$, and not $\bf H$, occurs in the interaction Lagrangian in Eq.~(\ref{1}).  Actually, $\bf B$ is the local spacetime average of the microscopical magnetic field usually called $\bf h$. Cf. Landau and Lifshitz \cite{landau84} for details.

\section{Electromagnetic energy-momentum tensor in a dielectric environment}

We now consider the conservation equations for energy and momentum for our open system, assuming as above that  $\varepsilon$ and permeability  $\mu$ are constants.

When analyzing energy-momentum problems  it is  convenient to start from the Poynting vector,
\begin{equation}
{\bf S}= {\bf E\times H}. \label{15}
\end{equation}
In ordinary electrodynamics when a wave falls  from  vacuum normally onto a dielectric surface, the same expression for $\bf S$ has to  hold in the interior also, as    the field is unable to do  work when passing a dielectric surface  at rest. Now from the  generalized Maxwell's equations we  calculate the energy conservation equation
\begin{equation}
{\bf \nabla \cdot S} + \dot{W}= -{\bf E \cdot J}-g_{a\gamma\gamma}{ \bf (E\cdot B)}\dot{a}, \label{16}
\end{equation}
where
\begin{equation}
W = \frac{1}{2}( \bf {E\cdot D} + {\bf H\cdot B} ) \label{17}
\end{equation}
is the electromagnetic energy density. There is thus in general an exchange of electromagnetic energy with the axion "medium", if $\bf E$ and $\bf B$ are different from zero and $a(t)$ is time-varying, even if ${\bf J}=0.$

Consider next the balance equation for electromagnetic momentum. This is a nontrivial point as one has to decide upon what is the correct expression for the momentum density $\bf g$. From a fundamental viewpoint it would appear as correct to employ Planck's principle of inertia of energy, saying that
 $ {\bf g=S}/c^2 $ (in physical units). That would mean adopting the expression known as Abraham's momentum density in the electrodynamics of media (here given a superscript A),
\begin{equation}
{\bf g}^{\rm A}= {\bf E\times H}. \label{18}
\end{equation}
Now consider  the usual Maxwell stress tensor,
\begin{equation}
T_{ik}= E_iD_k+H_iB_k-\frac{1}{2}\delta_{ik}({\bf E\cdot D+H\cdot B}). \label{19}
\end{equation}
This expression is common for the Abraham and Minkowski alternatives, $T_{ik}^{\rm A} = T_{ik}^{\rm M} \equiv T_{ik}$. In this way the momentum conservation equation takes the form
\begin{equation}
\partial_kT_{ik}-\dot{g}_i^{\rm A}= f_i^{\rm A},
\end{equation}
where $f_i^{\rm A}$ are the components of Abraham's force density
\begin{equation}
{\bf f}^{\rm A}= \rho {\bf E} + ({\bf J\times B}) + (\varepsilon \mu -1)\frac{\partial}{\partial t}({\bf E\times H})
-g_{a\gamma\gamma} {\bf (E\cdot B)\nabla}a.         \label{20}
\end{equation}
\emph{}In ordinary electrodynamics ($g_{a\gamma\gamma}= 0$), this expression agrees with that given by Landau and Lifshitz (Eq.~(75.18) in \cite{landau84}, or in Chaichian {\it et al.},  pages 53-58 \cite{chaichian16}, electrostriction omitted.
 Related treatments can be found at various places, for instance in \cite{moller72,chaichian16,ramos15,brevik79}. Electrostriction can be omitted because it  does not contribute to the total force on the axion cloud.

When $g_{a\gamma\gamma}=0$ the  intricate point in Eq.~(\ref{20}) is the third term on the right hand side, conventionally called the Abraham term.   It has experimentally turned up only in a few experiments, mainly at low frequencies where the mechanical oscillations of a test body  are directly detectable. Especially this is so in the classic Walker-Lahoz experiment from 1975 \cite{walker75,walker75a} (cf. also the recent considerations in Ref.~\cite{brevik22}), which tested the oscillations of a high-permittivity disk acting as a torsional pendulum. If we go to radiation pressure experiments in optics,  the action of the Abraham force will however  fluctuate out. It is therefore mathematically simpler, and in accordance with all observational experience  in optics, to include the Abraham momentum (physically, a mechanical accompanying momentum) in the effective field momentum. Therewith, the momentum density becomes simply the Minkowski momentum ${\bf g}^{\rm M}$, given by
\begin{equation}
{\bf g}^{\rm M}= {\bf D\times B}. \label{21}
\end{equation}
(More extensive discussions on this point are given in Refs.~\cite{brevik79,brevik18,brevik21}.)

Thus the momentum conservation equation becomes in the Minkowski case
\begin{equation}
\partial_kT_{ik}-\dot{g}_i^{\rm M}= f_i^{\rm M},
\end{equation}
where
\begin{equation}
{\bf f}^{\rm M}= \rho {\bf E}+({\bf J\times B})
-g_{a\gamma\gamma}  {\bf (E\cdot B)} {\bf \nabla} a.        \label{22}
\end{equation}
We can now construct a relativistically covariant form for the energy-momentum balance. Introduce the Minkowski energy-momentum tensor
\begin{equation}
S_\mu^{{\rm M}\nu}  = F_{\mu\alpha}H^{\nu\alpha}-\frac{1}{4}g_\mu^\nu F_{\alpha\beta}H^{\alpha\beta}, \label{23}
\end{equation}
which has the same form in all inertial frames. Then the conservation equations for electromagnetic energy and momentum can  be written as
\begin{equation}
-\partial_\nu S_\mu^{{\rm M}\nu}  = f_\mu ^{\rm M}, \label{24}
\end{equation}
where $f_\mu^{\rm M} = (f_0, {\bf f}^{\rm M})$ is the four-force density. In the rest system, ${\bf f}^{\rm M}$ is given by Eq.~(\ref{22}), whereas the zeroth component is
\begin{equation}
f_0= {\bf E}\cdot {\bf J}+g_{a\gamma\gamma} {\bf (E\cdot B)}\dot{a}. \label{25}
\end{equation}
We have written this component simply as $f_0$,  as it is the same in both formulations, $f_0^{\rm M}= f_0^{\rm A} \equiv f_0$.

We make a few remarks.

\noindent 1. Let us first apply the  expression (\ref{25}) for $f_0$ to the following  specific case, which is of special interest. Consider the situation analyzed  by Millar {\it et al.} \cite{millar17}: a dielectric surface $x=0$ divides space into two regions, one left region with refractive index $n_1$ and one right one with refractive index $n_2$. There is a strong magnetic field ${\bf B}_0$ in the $z$ direction parallel to the surface.  Because of the boundary condition ${\bf E}_\parallel =0$ at $x=0$ there will according to Ref.~\cite{millar17}  be generated two electromagnetic waves, one going to the left with Poynting vector ${\bf S}_1$, and one going to the right with Poynting vector ${\bf S}_2$. We will consider only the mean of the magnitude of the Poynting vector directed  to the right, called $S_2$,
\begin{equation}
{S}_2 = \frac{E_0^2}{2n_2^2}\left( \frac{1}{n_2}-\frac{1}{n_1} \right) \left[ 2\sin^2\left( \frac{k_2 x}{2}\right) -\frac{n_2}{n_1}\right],\quad k_2= n_2\omega,\label{25a}
\end{equation}
($\mu=1$ is assumed). $E_0$ is as defined in Eq.~(\ref{definition}). The expression (\ref{25a}) has the noteworthy property that it varies with distance $x$. Even the  one-dimensional gradient   $dS_2/dx $ varies with $x$. One may ask: what is the corresponding expression for $f_0$?

To see this, assume that the emitted photon field $ {\bf E}_{2\gamma}$ is polarized in the $z$ direction. From Ref.~\cite{millar17},
\begin{equation}
E_{2\gamma}= \frac{E_0}{n_2}\left( \frac{1}{n_2}-\frac{1}{n_1}\right) \cos(k_2x-\omega t),
\end{equation}
while the axion-generated field ${\bf E}_{2a}$ is given by Eq.~(\ref{15a}) with $\varepsilon = n_2^2$. Thus, we obtain for the time average
\begin{equation}
f_0= g_{a\gamma\gamma} \overline{(E_{2 \gamma}+E_{2a})B_0\dot{a}}=0.
\end{equation}
The first term averages out for arbitrary optical frequencies $\omega$, while the second term always averages out when $a=a_0\sin \omega_at$.  We can conclude that in the present  case the zeroth component $f_0$ of the four-force, usually associated with dissipative processes, does not on a time average take place in the pulsating interchange between the photon and the axion field energies.  This is as we might expect physically. What is more nontrivial, is that the time average  of ${\bf \nabla \cdot S} $ depends on $x$  under stationary conditions. With respect to the general energy conservation equation
\begin{equation}
{\bf{\nabla \cdot S}} + \dot{W} = -f_0,
\end{equation}
the behavior seems thus complicated. We  do not go into further detail here.

\noindent 2. The conservation equations (\ref{24}) can be written in a covariant form. The left hand side is covariant already, while the right hand side can be made covariant by constructing the four-force density $f_\mu^{\rm M}$ such that its components agree with the expressions (\ref{22}) and (\ref{25}) when the medium is at rest. In this sense the procedure is similar to that encountered with the Abraham covariant theory in ordinary electrodynamics. Cf. also the extensive discussions in Refs.~\cite{nesterenko16}.

\noindent 3. In astrophysical contexts one usually assumes  ${\bf \nabla}a=0$. According to Eq.~(\ref{22}) the force density is then zero when $\rho= {\bf J}=0$. At first sight this is  surprising, when comparing with the main idea behind  haloscopes which is to test the disturbance of a mechanical system acted upon by axions:  how can physical effects be measured if the force is zero? Part of the explanation may be that  the theory rests upon approximations, in particular,  neglect of dissipation. The plasma dispersion relation is
\begin{equation}
\varepsilon(\omega_a)=1+\frac{\omega_p^2}{\omega_0^2-\omega_a^2-i\gamma \omega_a}, \label{27}
\end{equation}
where $\omega_p$ is the plasma frequency and $\omega_0$  the eigenfrequency in the cylinder. Ordinarily, $\omega_0 \gg \omega_a$, but the idea behind the haloscope is effectively to dilute the cylindric system so as to let $\omega_0$ approach $\omega_a$. Therewith the magnitude of $\varepsilon$ increases. The expression (\ref{27}),  in itself independent of the magnitude of $\varepsilon$ as long as it is real, will have to change within the dissipative region when $\omega_0 \approx \omega_a$ and $\varepsilon$ becomes complex. We have to conclude that the  action of a haloscope rests on dissipation, and is independent of  the magnitude of the permittivity. A detailed calculation of axion electrodynamic effects is given by Kim {\it et al.} \cite{kim19}; cf. also the related Ref.~\cite{brevik22a}.

\section{Application 1:  Space-dependent axions}

We assume hereafter  that the usual electromagnetic charges and currents vanish, $\rho= {\bf J}=0$, and assume in the present section that $\dot{a}=0$ so that $a=a({\bf r})$ is a function of position only. This assumption is counter to the usual assumption that $a$ should be a function of time only. To some extent we thus introduce the space-dependent axion model as a theoretical model, motivated by the fact that it  often  leads to interesting mathematical results. Moreover, especially in the early universe such a model may be of physical interest also, since the Peccei-Quinn scalar field has been suggested  vary in space (typically as a tanh function of the spatial coordinate), leading in turn to space variations of the axion field also. In addition, in  galactic haloes the axions may  behave differently from what they do in the open space. The space-dependent model has been considered before; cf., for instance, Refs.~\cite{fukushima19} and \cite{brevik21a}.

We now allow for both magnetic and electric strong applied fields. The generalized Maxwell equations (\ref{5})-(\ref{9}) reduce to
\begin{equation}
{\bf \nabla \cdot D}= -g_{a\gamma\gamma}{\bf B\cdot \nabla}a, \label{28}
\end{equation}
\begin{equation}
{\bf \nabla \times H}= g_{a\gamma\gamma}{\bf \nabla}a\times{\bf E},   \label{29}
\end{equation}
\begin{equation}
{\bf \nabla \cdot B} = 0, \label{30}
\end{equation}
\begin{equation}
{\bf \nabla \times E}=0.    \label{31}
\end{equation}
It is natural to define $\rho_a$ as the axion electric charge density and ${\bf J}_a$ as the axion current density,
\begin{equation}
\rho_a = -g_{a\gamma\gamma}{\bf B\cdot \nabla}a, \label{32}
\end{equation}
\begin{equation}
{\bf J}_a =  g_{a\gamma\gamma}{\bf \nabla}a\times{\bf E}. \label{33}
\end{equation}
We will apply these equations to the situation where there is one homogeneous material plate present, of infinite extent in the horizontal $x$ and $y$  directions, extending in the vertical direction from $z=0$ to $z=L$. The material constants  $\varepsilon$ and $\mu$ in the plate are as before real constants. On the outside, there is a vacuum.  Assume that a strong magnetic field $B_0$ is applied in the $z$ direction, and a strong electric field $E_0$ applied in the $x$ direction,
\begin{equation}
{\bf B}_0 = B_0{\hat {\bf z}}, \quad {\bf E}_0= E_0{\hat{\bf x}}. \label{34}
\end{equation}
Axion-generated charges and currents are expected to be much smaller than those related to $B_0$ and $E_0$. We can put ${\bf B}= {\bf B}_0 + {\bf B}_a, \, {\bf E} = {\bf E}_0 + {\bf E}_a$, where the perturbations ${\bf B}_a$ and ${\bf E}_a$ are small. Thus we can replace the second order terms ${\bf B}_a \cdot {\bf \nabla}a$ and ${\bf \nabla}a \times {\bf E}_a$ by zero, whereby we obtain the reduced expressions
\begin{equation}
\rho_a = -g_{a\gamma\gamma}{\bf \nabla}\cdot (a{\bf B}_0), \label{35}
\end{equation}
\begin{equation}
{\bf J}_a= -g_{a\gamma\gamma}{\bf \nabla}\times(a{\bf E}_0). \label{36}
\end{equation}
Assume for definiteness that the axion field has a constant gradient in the $z$ direction inside the plate,
\begin{equation}
a(z) = \alpha z, \quad 0<z<L, \label{37}
\end{equation}
where $\alpha >0$ is constant, not necessarily small. It is convenient to express $\alpha$ as $\alpha = a_0/L$, where $a_0$ is the maximum value at $z=L.$ In the outside regions,  we assume $a(z)=0$.

Integrating (\ref{35}) across the boundary $z=L$ we obtain the surface charge density, called $\sigma_a$,
\begin{equation}
\sigma_a = g_{a\gamma\gamma}a_0B_0, \label{38}
\end{equation}
whereas in the interior region the volume charge density is
\begin{equation}
\rho_a= -g_{a\gamma\gamma}\alpha B_0. \label{39}
\end{equation}
Thus, the total charge $\rho_aL$ in the  interior region per unit surface is seen to balance the surface charge density,
\begin{equation}
\rho_aL +\sigma_a =0. \label{40}
\end{equation}
We see that after  imposition of the strong external fields $B_0$ and $E_0$,  the  plate remains electrically neutral. This is as we would expect physically: the appearance of these fields does not involve a supply of electric charge.

Correspondingly, for the current densities ${\bf J}_a$ we obtain first from (\ref{36}) a surface current density, called ${\bf K}_a$, as
\begin{equation}
{\bf K}_a = -g_{a\gamma\gamma}a_0 E_0 \hat{\bf y},\label{41}
\end{equation}
while in the interior
\begin{equation}
{\bf J}_a= g_{a\gamma\gamma}\alpha E_0 \hat{\bf y}. \label{42}
\end{equation}
Thus, analogously to above, Eq.~(\ref{40}), we see that the total current in the interior per unit surface, ${\bf J}_aL$, just balances the contribution from the surface,
\begin{equation}
{\bf J}_aL+{\bf K}_a= 0. \label{43}
\end{equation}
There is no net current in the $y$ direction. Note that $\varepsilon$ and $\mu$ do not appear in these expressions.

Consider next the axionic magnetic  and electric fields.  From Maxwell's equations (\ref{28}) and (\ref{29}) we obtain, when taking into account the smallness of these fields such as above,
\begin{equation}
{\bf H}_a= g_{a\gamma\gamma}\alpha z {\bf E}_0, \label{44}
\end{equation}
\begin{equation}
{\bf D}_a= -g_{a\gamma\gamma}\alpha z {\bf B}_0. \label{45}
\end{equation}
The induced magnetic and electric fields are thus respectively horizontal and vertical, increasing linearly in the $z$ direction. At the bottom of the plate, $z=0$, the induced fields vanish.

As ${\bf B}_0$ and ${\bf E}_0$ are orthogonal, it follows from Eqs.~(\ref{20}) and (\ref{22}) that the force density in the interior vanishes,
\begin{equation}
{\bf f}^{\rm A} =  {\bf f}^{\rm M}=0. \label{45a}
\end{equation}

So far, we have considered the interior Lorentz force. What about the force ${\bf F}_{\rm surface}$ on the  layer $z=L$? That force is actually zero, as one can see from a direct calculation using (\ref{38}) and (\ref{41}),
\begin{equation}
{\bf F}_{\rm surface}= \sigma_a{\bf E}_0 +{\bf K}_a\times {\bf B}_0 = 0. \label{49}
\end{equation}
There is thus in this case no electromagnetic force on the plate due to the interaction with axions.

The fourth component $f_0$ of the four-force density, Eq.~(\ref{25}), is also  zero, as it must be in a static situation.

We will now turn to the case where the axion field depends on time only.

\section{Application 2: Time-dependent axions}

Assume  that the axion field is  spatially uniform, but depends on  time, $a=a(t)$. We assume hereafter vacuum surroundings, so that $\varepsilon = \mu = 1$. Maxwell's equations become, when $\rho = {\bf J} =0$ as before,
\begin{equation}
{\bf \nabla \cdot E}= 0, \label{50}
\end{equation}
\begin{equation}
{\bf \nabla \times H}= \dot{\bf E}+g_{a\gamma\gamma}\dot{a}{\bf H}, \label{51}
\end{equation}
\begin{equation}
{\bf \nabla \cdot H}=0, \label{52}
\end{equation}
\begin{equation}
{\bf \nabla \times E} = -\dot{\bf H}, \label{53}
\end{equation}
while  the field equations (\ref{12}) and (\ref{13}) reduce to
\begin{equation}
\nabla^2 {\bf E}- \ddot{\bf E}=     g_{a\gamma\gamma} \dot{a}\dot{\bf B},  \label{54}
\end{equation}
\begin{equation}
\nabla^2 {\bf H}- \ddot{\bf H} = -g_{a\gamma\gamma} \dot{a}{\bf \nabla \times H}. \label{55}
\end{equation}
Our use of Eq.~(\ref{12}) instead of Eq.~(\ref{10}) relies upon the fact that   the axion-generated field ${\bf E}_a$ has been split off. We make  no assumption about the relative magnitudes of the optical frequency $\omega$ and the axion frequency $\omega_a$.

Assume now  that there is a wide region in the outer space where $a(t)$ has a simple sinusoidal variation,
\begin{equation}
a(t)= a_0\sin \omega_a t. \label{56}
\end{equation}
Since the axion velocity is most likely low, $v \sim   10^{-3}c$, we can put $\omega_a$ equal to the axion mass $m_a$. The value of this mass is not very well known, but we will take $m_a = 10~\mu$eV as a reasonable mean value. Thus in  physical units,  $\omega_a= 1.52 \times 10^{10}~$rad/s, which
corresponds to an oscillation wavelength of $\lambda_a=2\pi/\omega_a= 12.4$ cm. The axions are often associated with dark matter, whose energy density is estimated to be
$\rho_{\rm DM}= 0.35~$GeV/cm$^3$
\cite{qingdong19,read14}
\begin{equation}
\rho_a= \frac{1}{2}m_a^2a_0^2. \label{57}
\end{equation}
 We take the initial shape of an electromagnetic wave emitted from the Earth to be as
 \begin{equation}
 {\bf A}_0(x,t)=  c\,e^{-x^2/2D^2}\cos k_0x \,\hat{\bf y}, \label{58}
 \end{equation}
 where ${\bf A}_0$ is the vector potential, $c$ is the wave amplitude, and $k_0$ is the wave number. Writing the time-dependent wave as a Fourier integral,
 \begin{equation}
 {\bf A}_0(x.t)= \frac{1}{2} \int_{-\infty}^\infty \left[{\bf A}_0(k)e^{i(kx-\omega t)} + {\bf A}_0^* (k)e^{-i(kx-\omega t)} \right] dk, \label{59}
 \end{equation}
 we obtain by inversion
 \begin{equation}
 {\bf A}_0(k)= \frac{1}{2\pi}\int_{-\infty}^\infty e^{-ikx}\left[ {\bf A}_0(x,0)+\frac{i}{\omega}\frac{\partial {\bf A}_0}{\partial t}(x,0)\right]dx. \label{60}
 \end{equation}
 With the convenient assumption
 \begin{equation}
 \frac{\partial {\bf A}_0}{\partial t}(x,0)=0 \label{61}
 \end{equation}
 we then get
 \begin{equation}
 {\bf A}_0(k)= {\bf A}_0^+(k)+ {\bf A}_0^- (k), \label{62}
 \end{equation}
 where
 \begin{equation}
 {\bf A}_0^+(k)= \frac{cD}{2\sqrt{2\pi}}\exp\left[ -\frac{1}{2}D^2(k-k_0)^2\right]\, \hat{\bf y}. \label{62}
 \end{equation}
The expression for ${\bf A}_0^-(k)$ is the same, only with the substitution $k-k_0 \rightarrow k+k_0$. The right-moving wave, the part that we will keep in the following, is thus
\begin{equation}
{\bf A}_0^+(x,t)= \int_0^\infty {\bf A}_0^+(k)\cos(kx-\omega t)dk,\quad \omega = k >0. \label{63}
\end{equation}
We omit henceforth the superscript plus. The electric and magnetic fields of the incident wave are
\begin{equation}
{\bf E}_0(x,t)= -{\dot{\bf A}}_ 0(x,t),
 \quad {\bf H}_0(x,t)= {\bf \nabla \times A}_0(x,t). \label{64}
\end{equation}
Consider now the force from the incident wave on the axions. As $\rho = {\bf J}=0$ we  see from (\ref{20}) or (\ref{22}) that ${\bf f}=0$. This result might appear surprising, but is related to our assumption about a homogeneous axion cloud. The effect is analogous to that experienced in ordinary electrodynamics in a medium: the radiation force in the interior homogeneous region is zero, while the force in the boundary layer is a gradient force, ${\bf f}= -\frac{1}{2}E^2{\bf \nabla}\varepsilon.$

The dissipation component $f_0$ in (\ref{25}) is also zero, since the incident fields ${\bf E}_0$ and ${\bf H}_0$ are orthogonal. The axion-generated electric current $g_{a\gamma\gamma}\dot{a}{\bf H}_0$ as given by Eq.~(\ref{51}), is uninfluenced by the transverse field ${\bf E}_0$.

So, in order to calculate the resulting fields in the interacting wave-axion situation, we  go back to the field equations (\ref{54}) and (\ref{55}). This is actually the main idea also followed in Refs.~\cite{sikivie03} and \cite{arza19}. It is convenient first to rewrite (\ref{55})
as
\begin{equation}
\nabla^2{\bf A}-\ddot{\bf A}= -g_{a\gamma\gamma}\dot{a}{\bf \nabla \times A}, \label{65}
\end{equation}
in which we can make use of the expression (\ref{63}) directly, letting ${\bf A} \rightarrow {\bf A}_0$ on the right hand side.

Neglecting $\nabla^2{\bf A}$ on the left hand side and making use of the expressions (\ref{63}) and (\ref{64}) we get
\begin{equation}
\ddot{\bf A}(x,t)=      -g_{a\gamma\gamma}a_0\omega_a\int_0^\infty A_0(k)k\sin(kx-\omega t)\cos \omega_at\,dk \, \hat{\bf z}, \label{66}
\end{equation}
still with $\omega= k$.

We write the trigonometric product as a sum of two terms, and extract the term that leads to resonance. Going over to a complex representation,
\begin{equation}
\ddot{\bf A}(x,t)= \frac{1}{2}g_{a\gamma\gamma}a_0\omega_a{\rm Im}\int_0^\infty A_0(k)ke^{i(kx-\omega t+\omega_at)}\, \hat{\bf z}. \label{67}
\end{equation}
Defining ${\bf A}(k,t)$ via
\begin{equation}
{\bf A}(x,t)= {\rm Im}\int_0^\infty e^{ikx}{\bf A}(k,t)dk, \label{67}
\end{equation}
we can thus write
\begin{equation}
\ddot{\bf A}(k,t)=-\frac{1}{2}g_{a\gamma\gamma}a_0\omega_aA_0(k)ke^{-i(\omega-\omega_a)t}\, \hat{\bf z}. \label{68}
\end{equation}
Defining
\begin{equation}
{\bf {\cal{A}}}(k,t)={\bf A}(k,t)e^{-i\omega t}, \label{69}
\end{equation}
we calculate
\begin{equation}
\ddot{\bf A}(k,t)= \ddot{{\bf {\cal{A}}}}(k,t) e^{i\omega t} +2i\omega \dot{{\bf {\cal{A}}}}(k,t) e^{i\omega t}- \omega^2 {\bf {\cal{A}}}(k,t) e^{i\omega t}, \label{70}
\end{equation}
and keep only the resonance producing term containing $ \dot{{\bf {\cal{A}}}}(k,t)$. Then,
\begin{equation}
\dot{{\bf {\cal{A}}}}(k,t)  = \frac{i}{4}g_{a\gamma\gamma}a_0\omega_aA_0(k)e^{-i(2\omega-\omega_a)t}\, \hat{\bf z}. \label{71}
\end{equation}
After integration with respect to $t$, we can write this  vector potential component in the form
\begin{equation}
{\bf {\cal{A}}}(k,t) = -g_{a\gamma\gamma}a_0\omega_a\frac{cD}{8\sqrt{2\pi}}\exp\left[ -\frac{1}{2}D^2(k-k_0)^2\right]\frac{e^{-i(2\omega-\omega_a)t}}{2\omega-\omega_a}\, \hat{\bf z}. \label{72}
\end{equation}
The resonance occurs when $\omega= \omega/2$ as expected; an incoming frequency $2\omega$ can  split a resting axion into two components with equal mass $m_a=\omega_a$. Taking the imaginary part of the above expression, observing the limit
\begin{equation}
\lim \frac{\sin \alpha x}{\pi x}\big|_{\alpha \rightarrow \infty}= \delta(x), \label{73}
\end{equation}
and as result we have
\begin{equation}
{\rm{ Im}{{{\bf {\cal{A}}}}}}(k,t)  = g_{a\gamma\gamma}a_0\omega_a\frac{cD}{16}\sqrt{\frac{\pi}{2}}\exp\left[ -\frac{1}{2}D^2(k-k_0)^2\right]\delta(\omega-\frac{1}{2}\omega_a)\, \hat{\bf z}. \label{74}
\end{equation}
From this interaction term at resonance, one can calculate the axion echo. Such a calculation is basically given in Refs.~\cite{sikivie03} and \cite{arza19}, and will not be repeated here. The new element in the present calculation, is that it shows how the Gaussian profile in the emitted wave from the Earth influences the strength of the effect. To receive a  maximum echo, the center frequency $k_0=\omega_0$ in the pulse should be chosen equal to the resonance value of $k$, which is $\omega_a/2$.

Finally, from an energy-momentum point of view, we reemphasize the striking property of this kind of calculation that it is based on the field equations (\ref{54}) and (\ref{55}) instead of on the electromagnetic force density $\bf f$ given by Eqs.~(\ref{20}) or (\ref{22}).  This is quite uncommon in the ordinary  theory of  electromagnetic radiation forces.

\section*{Acknowledgment}

We are most grateful to Kimball A. Milton for several valuable remarks.

\end{document}